# Terahertz lattice and charge dynamics in ferroelectric semiconductor $Sn_xPb_{1-x}Te$

Y. Okamura[1*], H. Handa[1], R. Yoshimi[2], A. Tsukazaki[3],

K. S. Takahashi[2], M. Kawasaki[1,2], Y. Tokura[1,2,4] and Y. Takahashi[1,2*]

[1]*Department of Applied Physics and Quantum Phase Electronics Center, University of Tokyo, Tokyo 113-8656, Japan*

[2]*RIKEN Center for Emergent Matter Science (CEMS), Wako 351-0198, Japan*

[3]*Institute for Materials Research, Tohoku University, Sendai 980-8577, Japan*

[4]*Tokyo College, University of Tokyo, Tokyo 113-8656, Japan*

[*]To whom correspondence should be addressed (okamura@ap.t.u-tokyo.ac.jp, youtarou-takahashi@ap.t.u-tokyo.ac.jp)

**Abstract**

The symmetry breaking induced by the ferroelectric transition often triggers the emergence of topological electronic states such as Weyl fermions in ferroelectric-like metals/semimetals. Such strong coupling between the lattice deformation and electronic states is therefore essentially important for the control of novel topological phases. Here, we study the terahertz lattice and charge dynamics in ferroelectric semiconductor $Sn_xPb_{1-x}Te$ thin films hosting versatile topological phases by means of the terahertz time-domain spectroscopy. With lowering the temperature, the resonant frequency of transverse optical phonon shows the significant softening and upturn. This temperature anomaly of lattice dynamics directly indicates the displacive-type ferroelectric transition. The resulting phase diagram suggests the enhancement of ferroelectricity in the films due to compressive strain compared with the bulk crystals. The soft phonon induces the large DC and terahertz dielectric constant even in metallic state. Furthermore, we find that the Born effective charge of soft phonon mode is enhanced at around the compositions showing the band gap closing associated with the topological transition.

**INTRODUCTION**

Ferroelectrics, which spontaneously break the space inversion symmetry, have been investigated as one of the most fundamental phases of matter from the early stage of condensed matter physics[1]. So far, many ferroelectric materials and their unique physical properties such as the giant dielectric response, piezoelectric effect, and softening of transverse optical (TO) phonon have been extensively explored. The ferroelectrics now play a crucial role in today's device engineering such as capacitors and memories. Recently, the concept of ferroelectric-like metal, which defines a polar axis, expands the field of materials science. The ferroelectrics are usually considered to be characteristic of insulators, since the metallicity and ferroelectricity tend to be mutually exclusive with each other because the static internal electric field is screened by itinerant electrons. The progress in materials search has developed this new class of materials in this decade as exemplified by $LiOsO_3$[2], carrier-doped perovskite titanates[3,4] and transition metal dichalcogenides[5,6], while such a phase was originally predicted by a seminal theoretical work in 1960s[7]. These materials show intriguing phenomena related to both the charge and lattice dynamics such as ultrafast structural control by intense carrier acceleration[8] and enhanced thermopower near the boundary of the polar-to-nonpolar phase transition[9]. These findings suggest that the interplay between charge and lattice dynamics potentially provides the novel physical properties in ferroelectric-like metals. In particular, the softening of TO phonon in displacive-type ferroelectrics, which signals the structural instability and causes the enhanced dielectric response, is an intriguing target. However, the study on the dielectric properties of soft TO phonon in metals has been hardly addressed so far because the response of conduction electron overlaps with the low-energy dielectric response of TO phonon.

$Sn_xPb_{1-x}Te$ crystallizing in a rocksalt structure is the well-known IV-VI semiconductor exhibiting the ferroelectric instability (Fig. 1a). The electronic states of this compounds recently attract revived interests due to the appearance of topological phases depending on the composition $x$[10-12]. SnTe is known to be the topological crystalline insulator with polar distortion along <111> axes, which often shows the metallic conduction due to the Sn vacancy. The inelastic neutron and x-ray scattering measurements reveal the presence of the soft phonon at the Brillouin zone center below 4 meV, demonstrating the displacive-type ferroelectric transition[13,14]. It is suggested that the ferroelectric transition temperature strongly depends on the carrier concentration in SnTe[15]. On the other hand, PbTe is the trivial insulator with good thermoelectric performance under appropriate doping[16,17]. This material is also known as the incipient ferroelectric exhibiting the strong anharmonicity of optical phonon and local symmetry breaking, although the detailed nature is still controversial[18-21]. Inbetween the topological crystalline insulator for SnTe and trivial insulator for PbTe, the Weyl semimetallic state shows up with help of inversion symmetry breaking as suggested by the anomalous Hall and Nernst signals arising from the Weyl nodes in momentum space[22-26]. These studies indicate that the composition variation as well as the external pressure can induce the dramatic difference in the (topological) electronic and lattice states. The terahertz optical spectroscopy can directly measure the resonance of soft TO phonon and the conduction electron dynamics in the energy range of meV, providing an insight into the enhanced dielectric response associated with the structural instability in the topological semimetallic phase.

Here we investigate the terahertz lattice and charge dynamics of polar semiconductor $Sn_xPb_{1-x}Te$ thin films ($x$ = 0, 0.16, 0.30, 0.43, 0.63, 0.80, and 1) by means of terahertz spectroscopy. The fabrication of thin films enables the fine composition tuning, leading to the control of the ferroelectricity. In the terahertz conductivity spectra, we observe the clear phonon excitation for all the compositions as well as the electronic continuum band except for insulating PbTe ($x = 0$). The phonon excitations show the softening with decreasing temperature and then clear hardening for $0.16 \leq x \leq 0.80$. This observation indicates the robust ferroelectric transition under the presence of conduction electrons. The established phase diagram suggests the stabilization of the ferroelectricity compared to the bulk system most likely due to the strain from substrate. The low-energy soft phonon induces the enhanced DC dielectric constant. The enhancement of Born effective charge of TO phonon shows the clear correlation with the band gap closing.

## RESULTS

**Sample and characterization**

We used the high-quality (111)-oriented (Sn,Pb)Te epitaxial thin films with 2-nm thick SnTe buffer layer fabricated on InP(111)A substrate by molecular beam epitaxy (Fig. 1a; see also Methods). Figure 1b summarizes the temperature dependence of the resistivity of $Sn_xPb_{1-x}$Te thin films for $x = 0$, 0.30, 0.63, 0.80, and 1. The longitudinal resistivity $\rho$ exhibits the systematic composition $x$ dependence ranging over four orders of magnitude at 2 K; PbTe ($x = 0$) shows the semiconducting or insulating behavior while Sn-doped samples tend to show the metallic behavior. The Hall resistivity measurement in the previous work confirmed that Sn-substitution induces $p$-type carriers[25]. The appearance of polar axis along the out-of-plane [111] direction has been indicated by the optical second harmonic generation even for thin film PbTe, while PbTe in the bulk form is paraelectric (Fig. 1a). This suggests the stabilization of ferroelectricity in the present thin film form.

**Terahertz spectra**

We measured the terahertz conductivity and dielectric spectra by using the terahertz time-domain spectroscopy in the transmittance geometry (for details, see Methods). The terahertz spectra represent only the in-plane response for the normal incidence of the terahertz light. It should be emphasized that the terahertz measurement on thin films enables the quantitative evaluation of dielectric response of optical phonons, which directly reflects the emergence of ferroelectricity, even for the metallic samples. In bulk form, the metallic sample makes the measurement of terahertz spectra difficult owing to their reflectivity approaching unity.

The in-plane optical conductivity spectra Re $\sigma(\omega)$ for each composition are shown in Fig. 2. In PbTe (Fig. 2a), a peak of TO phonon exhibits clear softening with decreasing the temperature, indicating the development of lattice instability close to the ferroelectric phase transition. When substituting Pb with Sn, the continuum band due to the electronic excitation shows up in addition to the resonance of TO soft phonon. With increasing the Sn concentration $x$ or decreasing the temperature, the electronic continuum is increased in accord with the reduction in resistivity (Fig. 1b); Re $\sigma(\omega)$ in the lowermost energy well coincides with the DC conductivity obtained from the transport measurement as indicated by circles at zero energy (Figs. 2b-e). The phonon peak becomes unclear at the highly conductive regions, for example, SnTe at 5 K (purple curve in Fig. 2e).

To look into the soft phonon dynamics, we analyze the observed $\sigma(\omega)$ under the assumption that the terahertz response are composed of the electronic response $\sigma^{\text{ele}}(\omega)$ and phonon response $\sigma^{\text{ph}}(\omega)$. As an example, we show the fit to the terahertz spectra for Sn$_{0.30}$Pb$_{0.70}$Te at 5 K in Fig. 3 (purple curve in Fig. 2b) by using the following formula;

$$\sigma(\omega) = \sigma^{\text{ele}}(\omega) + \sigma^{\text{ph}}(\omega) = \sigma_0 + i\omega\varepsilon_0\left(1 - \varepsilon(\infty) - \varepsilon^{\text{ph}}(\omega)\right) \quad (1)$$

with

$$\varepsilon^{\text{ph}}(\omega) = \frac{S_{\text{ph}}\omega_T^2}{\varepsilon_0(\omega_T^2 - \omega^2 - i\omega\gamma)}. \quad (2)$$

$\varepsilon_0$, $\varepsilon(\infty)$, $\omega_T$, $S_{\text{ph}}$ and $\gamma$ are the dielectric constant of vacuum, dielectric constant in high-frequency limit, TO phonon energy, phonon's oscillator strength and damping constant, respectively. Here, we assume that $\sigma^{\text{ele}}(\omega)$ in the present energy window (1 - 8 meV) can be regarded as the lower-lying tail of the intraband/interband electronic excitation and approximated to be a frequency-independent constant value $\sigma_0$ because the DC conductivity is almost the same as Re $\sigma^{\text{ele}}(\omega)$ at $\hbar\omega = 8$ meV. The overall feature of the real and imaginary parts of the terahertz conductivity spectra is well reproduced by Eq. (1) (Figs. 3a,b). $\sigma_0$ almost coincides with the DC conductivity (green curves). The resonance structure due to the phonon excitation shows the resonance-type and dispersion-type spectra in Re $\sigma^{\text{ph}}(\omega)$ and Im $\sigma^{\text{ph}}(\omega)$, respectively (light blue curves). In a similar manner, other spectra shown in Fig. 2 can be decomposed into the resonant peak of soft TO phonon and flat continuum stemming from conduction electron except for the highly conductive regime, i.e., the Sn-rich compositions ($x = 0.80$, 1) at low temperatures. In these regions, the Drude-like behavior is indicated by the zero-energy peak in Re $\sigma(\omega)$; in case of SnTe at 5 K (purple curve in Fig. 2e), for example, Re $\sigma(\omega)$ shows the gradual decrease as a function of energy, where the scattering rate multiplied by $\hbar$ is estimated to be 23 meV.

The dielectric spectra obtained from the experiment and above fitting procedure are summarized in Figs. 4a-e. The real part of the dielectric spectra can be transformed from the optical conductivity spectra as,

$$\text{Re}\,\varepsilon(\omega) = 1 - \frac{\text{Im}\,\sigma(\omega)}{\varepsilon_0 \omega} = \text{Re}\,\varepsilon^{\text{ph}}(\omega) + \varepsilon(\infty) - \frac{\text{Im}\,\sigma^{\text{ele}}(\omega)}{\varepsilon_0 \omega}. \quad (3)$$

Here we omit $\varepsilon(\infty)$ and Im $\sigma^{\text{ele}}(\omega)$, because $\varepsilon(\infty)$ is as small as 40 for the present system[27,28], being negligible as compared to the $\varepsilon^{\text{ph}}(\omega)$, and Im $\sigma^{\text{ele}}(\omega)$ is nearly zero in most cases as discussed in Fig. 3b. In fact, the observed Re $\varepsilon(\omega)$ in Fig. 4a-e is well reproduced by Re $\varepsilon^{\text{ph}}(\omega)$ as indicated by the dotted curves. The dispersive structures with large amplitude manifest the enhanced dielectric response of soft TO phonon. The DC dielectric constant arising from the soft phonon $\varepsilon^{\text{ph}}(0)$ tends to be enhanced as the phonon softens at low temperatures (open circles, Figs. 4f-j); for example, it exceeds 2000 for the $x = 0.30$ compound (see open circles in Fig. 4g). This enhanced $\varepsilon(0)$ is consistent with the Lyddane-Sachs-Teller (LST) relation, $\varepsilon(0)/\varepsilon(\infty) = \omega_L^2/\omega_T^2$, in which the resonance energy of the longitudinal optical phonon, $\omega_L$, is as large as 17 meV[29,30].

## DISCUSSION

**Ferroelectric phase diagram**

The TO phonon energy $\omega_T$ obtained from fits with Eqs. (1) and (2) is plotted for each composition as a function of temperature (filled circles in Figs. 4f-j). With decreasing the temperature, TO phonon shows the significant softening for all the compositions down to the certain temperature, which is defined as $T_C$. The clear softening is observed even in metallic SnTe, being consistent with the previous inelastic x-ray scattering[14]; the screening effect less affects the emergence of ferroelectricity at least in the present carrier density (~ $6\times10^{20}$ cm$^{-3}$ for SnTe)[15]. The frequency lowering down to the $T_C$ can be well described by the Curie Weiss law, $\omega_T \propto \sqrt{T - T_C^*}$, as shown by dotted curves in Figs. 4f-j (see Fig. 4g for the definition of $T_C$ and $T_C^*$ as an example).

Below the $T_C$, the temperature dependence of the phonon energy shows the clear upturn in case of $x = 0.30$, $0.63$ and $0.80$. It is to be noted that the observed TO phonon has different polarization from the TO phonon causing the ferroelectric transition. The TO phonons are triply degenerated in the cubic paraelectric state and one of them is randomly frozen in the ferroelectric state in bulk crystal. In the thin film, the three modes are split into two in-plane polarized unfrozen modes observed here and one out-of-plane polarized frozen mode causing the ferroelectric transition at $T_C$ due to the epitaxial strain. It is well known that the soft phonon polarized along the polar direction shows the steep hardening below the ferroelectric transition temperature[31]. The unfrozen soft phonon modes polarized perpendicular to the polar direction show weaker hardening through the anharmonic coupling with other displacement modes allowed by the symmetry breaking in the Landau free energy[31,32]. Therefore, the upturn of the in-plane phonon energy observed in Figs. 4g-i can be regarded as the signature of the ferroelectric transition. In case of $x = 0$, while the temperature dependence of phonon energy shows the weaker anomaly, the deviation from the Curie Weiss law should be ascribed to the ferroelectric transition, in accord with the emergence of optical second harmonic generation enhanced at low temperatures[25].

In Fig. 5a, we show the ferroelectric phase diagram of $Sn_xPb_{1-x}Te$ thin film deduced from the terahertz spectroscopy. Both $T_C^*$ and $T_C$ tend to increase with increasing the Sn concentration $x$ as observed in the bulk system[15,33,35]. Meanwhile, the transition temperature for the present thin film system is much enhanced compared to the bulk system, demonstrating the stabilization of the ferroelectric phase. The lattice constant of thin film PbTe shrinks as compared to the bulk case,[25] indicating the compressive strain at least for Pb-rich regions. This should promote the elongation along the out-of-plane [111] direction, resulting in the enhancement of ferroelectric phase in particular for the Pb-rich regions.

**Enhanced Born effective charge**

We discuss the oscillator strength of the optical phonons in terms of Born effective charge. The dielectric response of TO phonon in Eq. (2) can be described by explicitly using the Born effective charge $Z^*$ as,

$$\varepsilon^{\text{ph}}(\omega) = \frac{N(Z^*e)^2}{M^*\varepsilon_0} \frac{1}{\omega_T^2 - \omega^2 - i\omega\gamma}, \qquad (4)$$

where $N$, $e$ and $M^*$ represent the oscillator density, elementary charge and reduced mass, respectively. The $Z^*$ for PbTe and SnTe are respectively estimated to be ~ 5.2 and ~ 5.3, both of which are much larger than the formal valence of Pb or Sn ion 2+. These enhanced $Z^*$ are roughly consistent with the first-principles calculation showing $Z^* = 6.16$ for PbTe and $Z^* = 8.26$ for SnTe (blue squares, Fig. 5b)[35].

Notably, the $Z^*$ shows the appreciable enhancement up to ~ 8 at around $x \simeq 0.3$ (Fig. 5b), at which the appearance of Weyl semimetal phase is suggested in the bulk system[23,26]. The enhanced mobility and second harmonic generation also imply the possible Weyl semimetallic state around $x \simeq 0.16$ in the present film[25]. The dielectric constant stemming from the higher-lying electronic states, which is effectively expressed by $\varepsilon(\infty)$, increases owing to the closing of bulk band gap near the Weyl semimetallic state. According to the LST relation, $Z^*$ is proportional to $\varepsilon(\infty)(\omega_L^2 - \omega_T^2)$, so that the increase of $Z^*$ can be ascribed to the $\varepsilon(\infty)$ enhanced by the band gap closing near the topological transition. The robust low-energy soft phonon produces the large dielectric response with assist of band gap closing in the present ferroelectric semiconductors. Note that the $\omega_L$ shows little composition dependence ($\omega_L \simeq 17$ meV)[29,30] and therefore phonon frequencies ($\omega_L^2 - \omega_T^2$) less affect $Z^*$.

In summary, we have investigated the terahertz response of the ferroelectric semiconductor $Sn_xPb_{1-x}Te$ thin film. We observe the coexistence of the clear soft phonon excitation and the electronic continuum band in the terahertz conductivity spectra. The TO phonon shows the softening and then slightly hardening irrespective of the change in the DC conductivity depending on the composition. This observation indicates the ferroelectric transition relatively robust against the presence of conduction electrons. The phase diagram obtained from the terahertz spectroscopy suggests the stabilization of the ferroelectric phase compared with the bulk crystals due to compressive strain characteristic of the thin films. It is found that the large DC dielectric constant induced by the soft TO phonon is further enhanced at around $x = 0.3$ owing to the band gap closing. Since the present soft phonon excitation is theoretically suggested to be strongly coupled to the electronic band at the L point of the Brillouin zone, where the band inversion occurs[21], the intense excitation of the soft phonon with a large dielectric response may open a new route for the ultrafast optical control of the topological electronic states including the Dirac surface state and Weyl semimetallic states.

**METHODS**

**Sample fabrication and characterization of transport properties**

We fabricated 40-nm thick (111)-oriented (Sn,Pb)Te epitaxial thin films on InP(111)A substrate by molecular beam epitaxy. The epi-ready substrate was annealed at 350 °C in a vacuum before the epitaxy and then set to 400 °C to grow the thin film. We inserted 2-nm thick SnTe buffer layers beneath the (Sn,Pb)Te layer, which helps to stabilize (111) orientation of (Sn,Pb)Te thin films. The growth temperature of the buffer layer was also 400 °C. We set the equivalent pressures of Sn and Te for the buffer layer at $P_{Sn} = 5.0 \times 10^{-6}$ Pa and $P_{Te} = 1 \times 10^{-4}$ Pa, respectively. For the (Sn,Pb)Te layer, we keep the sum of the equivalent beam pressures of Sn and Pb as $P_{Sn} + P_{Pb} = 1 \times 10^{-5}$ Pa, and change their ratio to change the Pb composition. We calibrated the actual Pb composition by inductively coupled plasma mass spectroscopy and defined it as $x$ in the main text. The calibrated $x$ value was larger than the nominal one by 0.1 ~ 0.2. The growth duration of the SnTe buffer layer and (Sn,Pb)Te layer was respectively 2 and 30 minutes regardless of $x$. We measured the transport properties using the standard four-terminal method by the Physical Properties Measurement System (PPMS, Quantum Design).

**Terahertz time-domain spectroscopy**

The laser pulses with a duration of 100 fs from a mode-locked Ti:sapphire laser were split into two paths to generate and detect the terahertz light by using the photoconductive antenna. The complex transmission spectrum $t(\omega)$ is obtained by comparing the Fourier transformation of the electric-field transmission through the (Sn,Pb)Te/InP and InP substrate. We then deduced the terahertz conductivity spectra $\sigma(\omega)$ from the following standard formula; $t(\omega) = (1+n_s)/(1+ n_s + Z_0 d \sigma(\omega))$, where $d$, $Z_0$, and $n_s$ are the thickness of the film, the vacuum impedance (377 Ω), and the refractive index of the InP substrate, respectively.

**DATA AVAILABILITY**

The data that support the plots of this study are available from the corresponding author upon reasonable request.

**ACKNOWLEDGEMENTS**

We thank J. Fujioka for fruitful discussion. This work is partly supported by JSPS/MEXT Grant-in-Aid for Scientific Research (No. 21H01796), and JST CREST (No. JPMJCR1874).

**AUTHOR CONTRIBUTIONS**

Y. Tokura and Y. Takahashi conceived the project. Y.O. and H.H. performed the terahertz measurement and analyzed the data. R.Y. fabricated the thin films under supervision of A.T., K.S.T., M.K. and Y. Tokura. Y.O., H.H., R.Y., Y. Tokura and Y. Takahashi discussed and interpreted the results with inputs from other authors. Y.O. and Y. Takahashi wrote the manuscript with the assistance of other authors.

**COMPETING INTERESTS**

The authors declare no competing financial interests.

**ADDITIONAL INFORMATION**

**Figures**

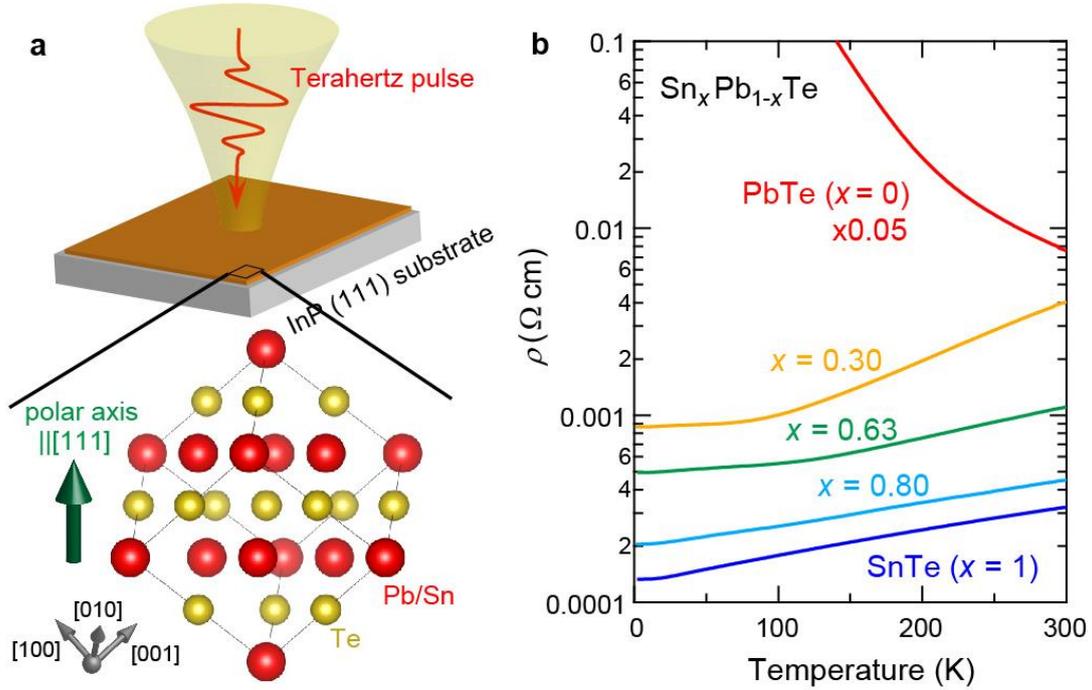

**Fig. 1   Crystal structure and transport property of $Sn_xPb_{1-x}Te$ thin films. a** Crystal structure of $Sn_xPb_{1-x}$Te. The thin film was fabricated on InP (111) substrate. The polar axis along the out-of-plane [111] direction is inferred from the second harmonic generation in the previous study[25]. The terahertz light is incident normal to the plane. **b** The temperature dependence of the resistivity for $Sn_xPb_{1-x}$Te ($x$ = 0, 0.30, 0.63, 0.80, 1).

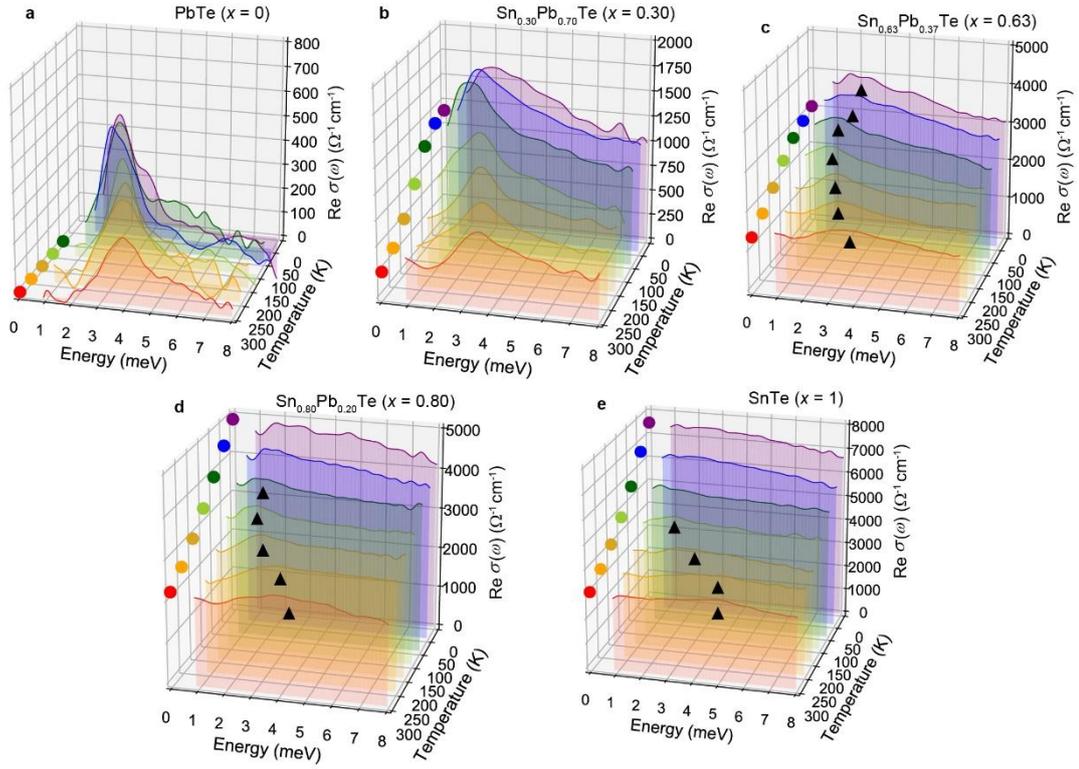

**Fig. 2 Terahertz conductivity spectra of $Sn_xPb_{1-x}Te$ thin films. a-e** Temperature dependence of the real part of the optical conductivity spectra Re $\sigma(\omega)$ for $Sn_xPb_{1-x}Te$ ($x$ = 0, 0.30, 0.63, 0.80, 1). The filled circles in zero-energy planes represent the DC conductivity obtained from the transport measurement. The data at 5 K (purple), 50 K (blue), 100 K (green), 150 K (yellow green), 200 K (ocher), 250 K (orange) and 300 K (red) are plotted. In **c**-**e**, the phonon excitations are indicated by black triangles for clarity.

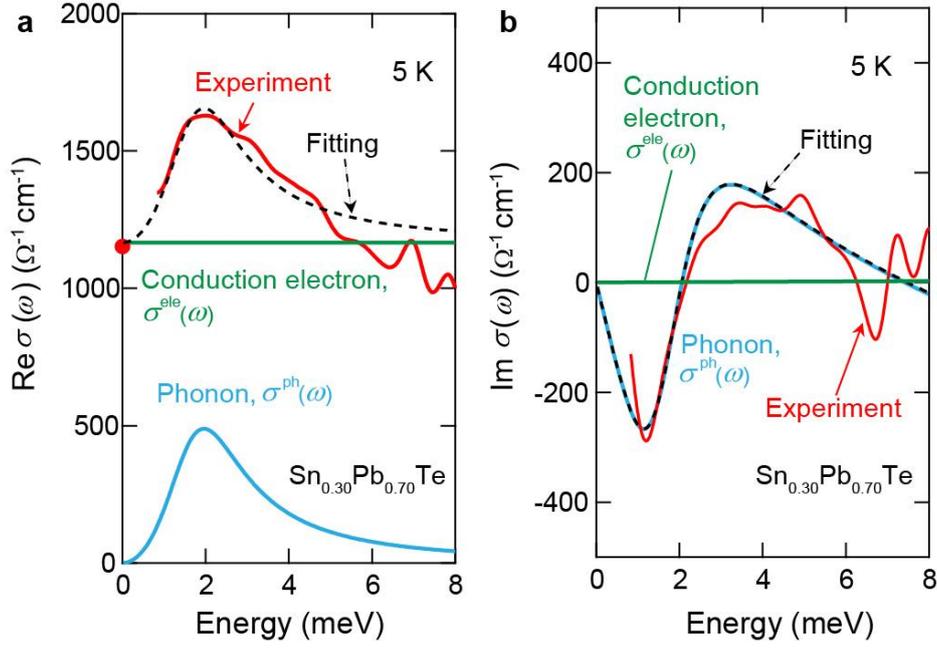

**Fig. 3 Decomposition of the terahertz conductivity spectra into the electronic and phonon responses.** Real (**a**) and imaginary (**b**) parts of $\sigma(\omega)$ at 5 K for $Sn_{0.30}Pb_{0.70}Te$. Red solid, green solid, light blue solid and black dashed curves represent the experimental spectrum, electronic continuum band $\sigma^{ele}(\omega)$, phonon response $\sigma^{ph}(\omega)$ and fitting spectrum, respectively. The red filled circle at zero energy at **a** denotes the DC conductivity obtained from the transport measurement.

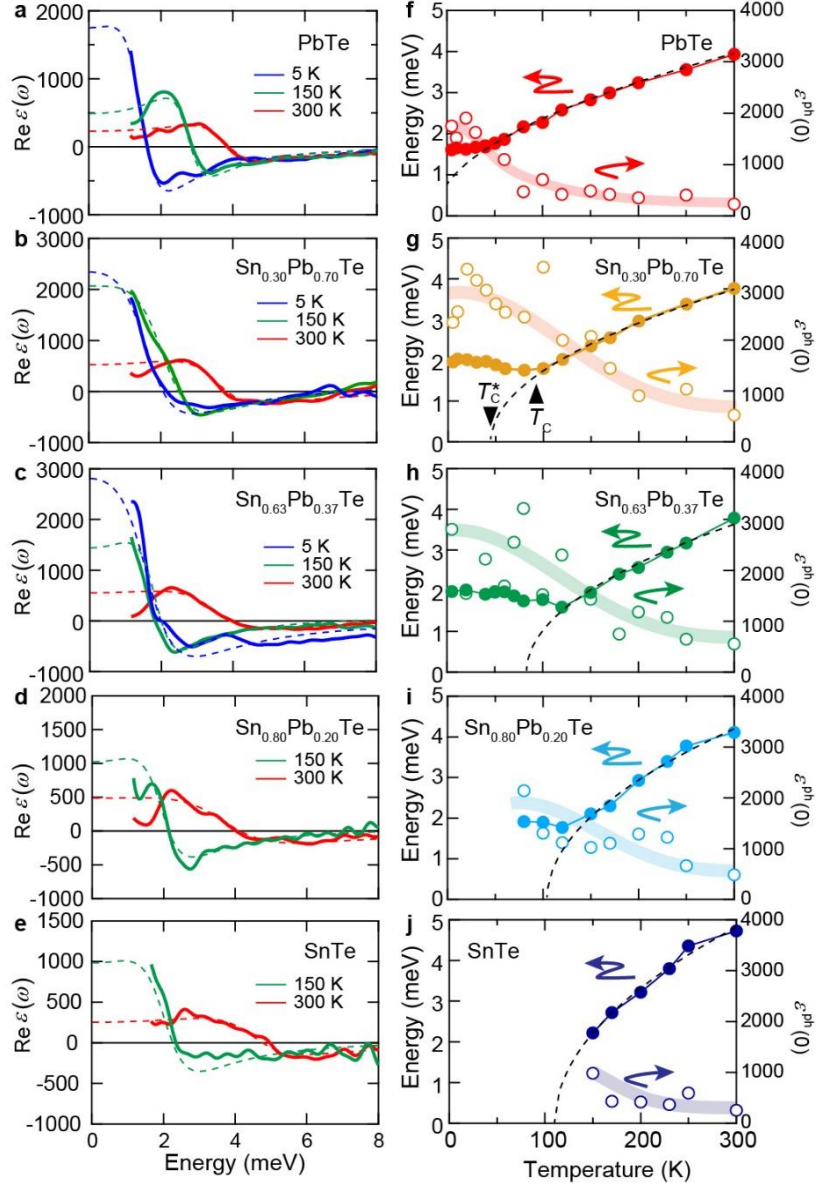

**Fig. 4 Soft phonon dynamics. a-e** Selected terahertz in-plane dielectric spectra Re $\varepsilon(\omega)$ for $Sn_xPb_{1-x}Te$ ($x$ = 0, 0.30, 0.63, 0.80 and 1). The dotted curves are fits with Re $\varepsilon^{ph}(\omega)$ in Eq. (2). **f-j** Temperature dependence of the soft TO phonon energy $\omega_T$ obtained from fits with Eqs. (1) and (2) (filled circles) and DC dielectric constant arising from the soft TO phonons Re $\varepsilon(\omega = 0)$ (open circles). The dotted curves represent the Curie Weiss law. The definitions of $T_C^*$ and $T_C$ are indicated in Fig. 4g as an example. The thick curves are the guide to the eyes.

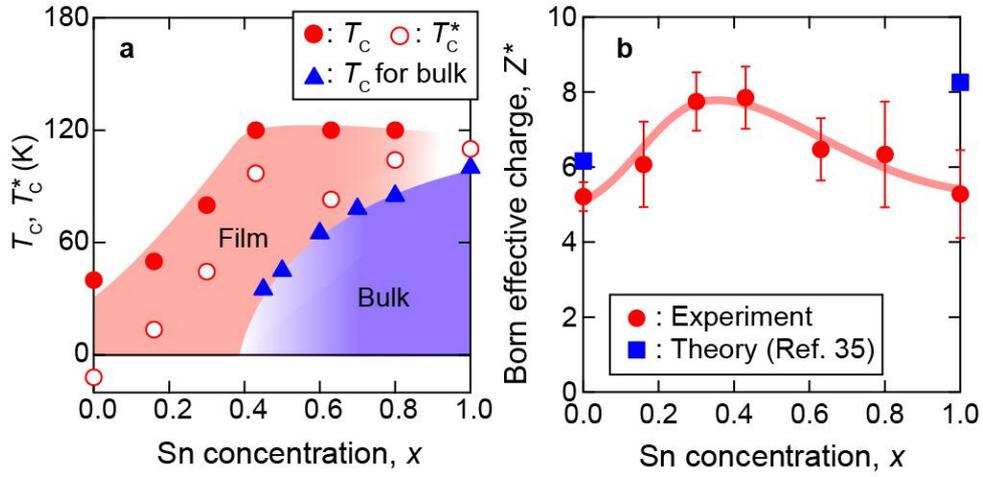

**Fig. 5 Ferroelectric phase diagram and enhancement of Born effective charge. a** Ferroelectric phase diagram determined from the soft phonon spectroscopy. The red filled and open circles respectively represent $T_C$ and $T_C^*$ (see main text for definition). The blue filled triangles are the transition temperature for the bulk system taken from refs. 15, 33, 34. While the $T_C$ depends on the carrier concentration at least for bulk SnTe, we plot the highest $T_C$ reported in ref. 15. **b** The $x$ dependence of the Born effective charge $Z^*$. The red circles and blue squares denote the experimental values and theoretical ones from the first-principles calculation[35], respectively. The vertical bars represent the thermally induced variation, which is defined as the standard deviation of the $Z^*$ calculated at each temperature. The red thick curve is the guide to the eyes.